\documentclass[runningheads]{llncs}
\usepackage{graphicx}
\usepackage{amssymb}
\usepackage{amsmath}
\usepackage{multirow}
\usepackage{booktabs,siunitx}
\usepackage{hyperref}
\DeclareMathOperator{\vect}{vec}
\begin{document}
\title{Hessian-based Similarity Metric for Multimodal Medical Image Registration}
%
%
\author{Mohammadreza Eskandari\inst{1,2} \and
Houssem-Eddine Gueziri\inst{2} \and
D. Louis Collins\inst{1,2,3}}
\authorrunning{M. Eskandari et al.}
%
\institute{Department of Biomedical Engineering, McGill University, Montreal, QC, Canada \email{mohammadreza.eskandari@mail.mcgill.ca}
\and
McConnell Brain Imaging Center, Montreal Neurological Institute and Hospital, Montreal, QC, Canada
\and
Department of Neurology and Neurosurgery, McGill University, Montreal, QC, Canada}
\maketitle              
\begin{abstract}
One of the fundamental elements of both traditional and certain deep learning medical image registration algorithms is measuring the similarity/dissimilarity between two images. In this work, we propose an analytical solution for measuring similarity between two different medical image modalities based on the Hessian of their intensities. First, assuming a functional dependence between the intensities of two perfectly corresponding patches, we investigate how their Hessians relate to each other. Secondly, we suggest a closed-form expression to quantify the deviation from this relationship, given arbitrary pairs of image patches. We propose a geometrical interpretation of the new similarity metric and an efficient implementation for registration. We demonstrate the robustness of the metric to intensity nonuniformities using synthetic bias fields. By integrating the new metric in an affine registration framework, we evaluate its performance for MRI and ultrasound registration in the context of image-guided neurosurgery using target registration error and computation time.

\end{abstract}
\section{Introduction}
Aligning images of different modalities is the key to combining functional and anatomical data from multiple sources in image-guided procedures. In this context, quantifying the similarity of images is an important yet challenging task due to different intensity distribution arising from distinct protocols and physical principles governing each modality. This already-complex problem becomes more complicated in the presence of noise and modality-dependent artifacts. A huge variety of mathematically proven, hand-crafted, and deep learning-based solutions have been proposed for many use cases, each having their own advantages and disadvantages.

One of the earliest successes in multimodal registration was achieved by the introduction of mutual information~\cite{ref9,ref20}. This measure is very general since it only assumes statistical dependence between the intensities of the images. To improve its performance for certain tasks, several variants have been proposed~\cite{ref8,ref15,ref16}. Soon after mutual information, the correlation ratio was introduced~\cite{ref17}, with the idea
of constraining the relationship between the intensities of the images. The correlation ratio assumes a functional dependence between image intensities and quantifies the compliance of this assumption. There exist some variants of the correlation ratio, however this method did not gain the popularity of mutual information.

Both mutual information and correlation ratio use image intensities directly as the feature space. Another research avenue in multimodal registration has been handcrafting features. Image gradients, local statistical information and local structures such as points, edges, contours, surfaces and volumes, are some commonly used~\cite{ref13}. Many studies have also tried to convert multimodal registration to unimodal registration by simulating one image from the other~\cite{ref1,ref22} or mapping both images into a third space~\cite{ref21}.

More recently, several deep learning methods have been proposed to overcome the challenges of measuring similarity between image pairs of distinct modalities~\cite{ref2,ref6,ref18}. The main idea behind these metrics is to use supervised learning to model similarity between registered pairs of patches. As a result, these methods can only perform well on the modalities they have been trained on and generalization to other modalities is highly dependent on data augmentation methods used during training.

One of the well-studied features for image registration and medical image processing in general is image gradient. The gradient can be a good descriptor of a neighborhood; its magnitude and direction describe how the intensities change in a small radius. A popular similarity metric for multimodal registration is the cosine squared of the angle between gradients of two images, which we will refer to as gradient orientation alignment~\cite{ref12}. This method was first intuitively proposed to incorporate local information~\cite{ref5,ref14}. Later, it was shown that maximizing mutual information on a small patch is equivalent to maximizing gradient orientation alignment~\cite{ref7}. The usefulness of gradient orientation alignment raises the question whether higher order derivatives can be beneficial in quantifying image similarities.

Despite the long-run progress in research on traditional similarity measures and transition to deep learning era, we believe there still exists certain areas to be explored for defining better similarity metrics based on mathematical properties of the images. Our focus is on the Hessian matrix, which contains information regarding the geometry of the image in a neighborhood. One of the most important applications of Hessian matrix in medical image processing is identifying vessel-like structures~\cite{ref4}. This is because a Hessian can approximate a neighborhood with an ellipsoid. Our approach in this paper is similar to that of correlation ratio~\cite{ref17}, since we constrain the relation between the two modalities with a functional dependence. We develop the mathematical formulation for using Hessians in measuring image similarity and test the performance of our proposed metric in synthetic and real use cases.
\section{Method}
\subsection{Defining the Hessian-based Similarity Metric}
\begin{figure}
\includegraphics[width=\textwidth]{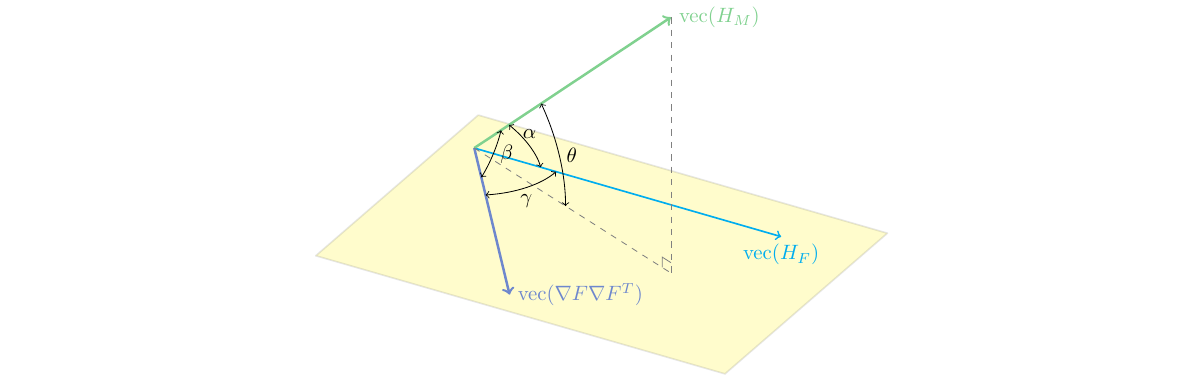}
\caption{Geometric interpretation of the Hessian-based similarity metric} \label{fig1}
\end{figure}

We define a metric for quantifying pointwise similarity between two medical images
of distinct modalities. We refer to the images as the fixed image and the moving image and denote them with \(F(\mathbf{x}), M(\mathbf{x}): \Omega \subset \mathbb{R}^d \rightarrow \mathbb{R}\), respectively, where \(d\) is the
dimension of the images. We start by deriving the relationship between the Hessians of two aligned patches centered at \(\mathbf{x_{0}}\), assuming a functional dependence between their intensities, i.e., there exists a function \(g: \mathbb{R} \rightarrow \mathbb{R}\) that maps the intensities of the fixed patch onto the corresponding intensities of the moving patch. The gradients and the Hessians of the images are represented by \(\nabla F\), \(\nabla M\), \(H_F\) and \(H_M\), respectively. For any \(\mathbf{\delta x}\), as long as \(\mathbf{x_{0} + \delta x}\) lies inside the patch of interest, we can formulate the functional dependence assumption as:
\begin{equation}
M(\mathbf{x_{0} + \delta x}) = g(F(\mathbf{x_{0} + \delta x})) \,.
\end{equation}
Writing the Taylor expansion of both sides up to the second order and equating terms
of the same order yields:
\begin{equation}
\nabla M(\mathbf{x_{0}}) = \frac{dg}{dF} \nabla F(\mathbf{x_{0}}) \,.
\end{equation}
\begin{equation}
H_M(\mathbf{x_{0}}) = \frac{dg}{dF} H_F(\mathbf{x_{0}}) + \frac{d^2g}{dF^2} \nabla F(\mathbf{x_{0}})\nabla F^T(\mathbf{x_{0}}) \,.
\end{equation}
Since the mapping between the intensities is not given, the derivatives of \(g\) are unknown. We rewrite the above equations by replacing the derivatives of \(g\) with scalar
variables \(\lambda\), \(\mu\) and \(\nu\). Note that these equations are valid for any point inside the patch
of interest, therefore for convenience of notation, we drop the dependence of the Hessians and the gradients on \(\mathbf{x}\).
\begin{equation} \label{eq:4}
\nabla M = \lambda \nabla F \,.
\end{equation}
\begin{equation} \label{eq:5}
H_M = \mu H_F + \nu \nabla F\nabla F^T \,.
\end{equation}
These two equations describe how the gradients and the Hessians of two patches relate to each other, subject to a functional dependence between their intensities. The first equation encapsulates \(d\) scalar equations and suggests that the gradients of two patches should be aligned or anti-aligned everywhere (depending on the sign of \(\lambda\)).
The second equation encapsulates \(d^2\) equations, only \((d^2+d)/2\) of which are unique due to symmetricity of matrices. It can be inferred from this equation that the Hessian of the moving patch can be decomposed into a linear combination of the Hessian and the gradient outer product of the fixed image. In other words, there exist two scalars, \(\mu\) and \(\nu\), such that they satisfy all encapsulated scalar equations at the same time. We use this equation as the starting point for defining our Hessian-based similarity metric. For arbitrary pairs of patches, there is no functional dependence between their intensities and hence, Eq.~\ref{eq:5} will not hold true. To quantify the violation of this equality, we define a normalized quadratic error:
\begin{equation}
E = \frac{||H_M - \mu H_F - \nu \nabla F \nabla F^T||^2}{||H_M||^2} \,,
\end{equation}
where \(||.||\) denotes the Frobenius norm. To evaluate this error regardless of the unknown scalars, we find \(\mu\) and \(\nu\) such that \(E\) will be minimized. We denote the optimal solution with \(\mu^\star\) and \(\nu^\star\) and the optimal value with \(E^\star\). To minimize \(E\), we compute its derivatives with respect to \(\mu\) and \(\nu\), equate the derivatives to zero, solve the resulting system of linear equations, obtain \(\mu^\star\) and \(\nu^\star\) and plug them into equation. \(E^\star\) is a measure of dissimilarity and is guaranteed to be bounded by zero and one. We define
the Hessian-based similarity metric as \(S=1-E^\star\). This metric can be represented by a closed-form expression:
\begin{equation}
\scriptsize
    S =\frac
{||\nabla F||^4 tr(H_M^T H_F)^2+||H_F||^2 (\nabla F^T H_M \nabla F)^2 -2tr(H_M^T H_F)(\nabla F^T H_M \nabla F)(\nabla F^T H_F \nabla F)}
{||H_M||^2(||\nabla F||^4||H_F||^2-(\nabla F^T H_F \nabla F)^2)} \,,
\end{equation}
where \(S=1\) implies perfect functional dependence between intensities and \(S=0\) implies perfect functional independence. As it was expected from Eq. \ref{eq:5}, the metric is assymetric in terms of fixed and moving image. Another way to express this metric is by vectorizing Eq.~\ref{eq:5}. We can formulate the metric in terms of \(\alpha\), \(\beta\) and \(\gamma\) which denote the angle between \(\vect(H_M)\) and \(\vect(H_F)\), the angle between \(\vect(H_M)\) and \(\vect(\nabla F \nabla F^T)\), and the angle between \(\vect(H_F)\) and \(\vect(\nabla F \nabla F^T)\), respectively.
\begin{equation}
S=\frac{\cos^2\alpha+\cos^2\beta-2\cos\alpha\cos\beta\cos\gamma}{\sin^2\gamma} \,.
\end{equation}
Additionally, the Hessian-based similarity metric has a clear geometric interpretation. It can be interpreted as the cosine squared of the angle between \(\vect(H_M)\) and its projection onto the plane spanned by \(\vect(H_F)\) and \(\vect(\nabla F \nabla F^T)\).  The angles between the vectorized matrices is depicted in Figure~\ref{fig1}. It must be taken into consideration that vectorized Hessian matrices can not be visualized in 3~dimensions and this figure is only intended for providing intuition regarding the defined metric.

We should mention that by following the above steps for Eq.~\ref{eq:4} instead of Eq.~\ref{eq:5}, the resulting similarity metric will be the cosine squared of the angle between the gradients of the two images. The derivation of this metric is included in supplementary material.

In practice, to compute the gradients and the Hessians of the images, we convolve them with first and second order Gaussian derivative kernels. The choice of standard deviation for these Gaussian kernels directly affects the behavior of the similarity metric and enables capturing structures of different scales. The standard deviation should be the same for both images and all components of the Gaussian kernels. By setting the standard deviation to a certain scale, structures of higher scale will not be captured and structures of lower scale will be smoothed out.

\subsection{Transforming Hessians}
A straightforward strategy for computing the Hessian-based similarity metric during the registration process involves first applying a deformation onto the moving image and then computing its Hessian. Since the cost of computing the Hessian is relatively high, our approach is to compute it only once at the preprocessing stage and to transform it according to each deformation.

Assuming a diffeomorphism \(\mathbf{P}(\mathbf{x}): \mathbb{R}^d \rightarrow \mathbb{R}^d\) that maps \(\mathbf{x}=(x_1, ..., x_d)\) from the original image onto \(\mathbf{u}=(u_1, ..., u_d)\) from the deformed image, each component of the Hessian in the transformed coordinate can be expressed in terms of the components of the original Hessian and gradient:
\begin{equation} \label{eq:9}
\frac{\partial^2 M}{\partial u_k \partial u_l}=\sum_{i=1}^{d} \frac{\partial M}{\partial x_i}\frac{\partial^2 x_i}{\partial u_k \partial u_l}+\sum_{i=1}^{d}\sum_{j=1}^{d} \frac{\partial^2 M}{\partial x_i \partial x_j} \frac{\partial x_i}{\partial u_k} \frac{\partial x_j}{\partial u_l} \,.
\end{equation}
Partial derivatives of \(\mathbf{x}\) with respect to \(\mathbf{u}\) are directly computable if the inverse transformation, \(\mathbf{P}^{-1} (\mathbf{u})\), is explicitly known. Eq.~\ref{eq:9}  implies that a nonlinear transformation must be \(C^2\) continuous, otherwise the transformed Hessian will be discontinuous. Note that when the transformation is linear, the first term on the right hand of Eq.~\ref{eq:9} will vanish and the transformed Hessian can be expressed as \(J_{\mathbf{P}} ^{-T} H_M J_{\mathbf{P}}^{-1}\), where \(J_{\mathbf{P}}\) is the Jacobian matrix of the transformation.
\subsection{Implementing the Metric in an Affine Registration Scheme}
We present a simple affine registration scheme to test the proposed similarity metric. We start by computing the Hessian of both images and the gradient of the fixed image. We sample \(N\) random voxels from the fixed image. The Hessian-based similarity metric is computed over the sampled voxels and their average is used as the similarity score for each deformation. No regularization term is used in this scheme. We use Differential Evolution~\cite{ref19} to find the optimal affine deformation. Affine deformations are defined by a translation vector and three matrices representing rotation, shear, and scaling. Shear and scaling are defined as an upper triangular matrix and a diagonal matrix, respectively. For each affine deformation, its inverse is applied to the location of the sampled voxels to find their corresponding points on the moving image. Hessian of the moving image is linearly interpolated over these points and transformed using the Jacobian of the deformation field. The optimization process terminates as soon as a termination condition is satisfied.
\begin{figure}
\includegraphics[trim={6cm 2cm 4cm 1cm},clip, width=\textwidth]{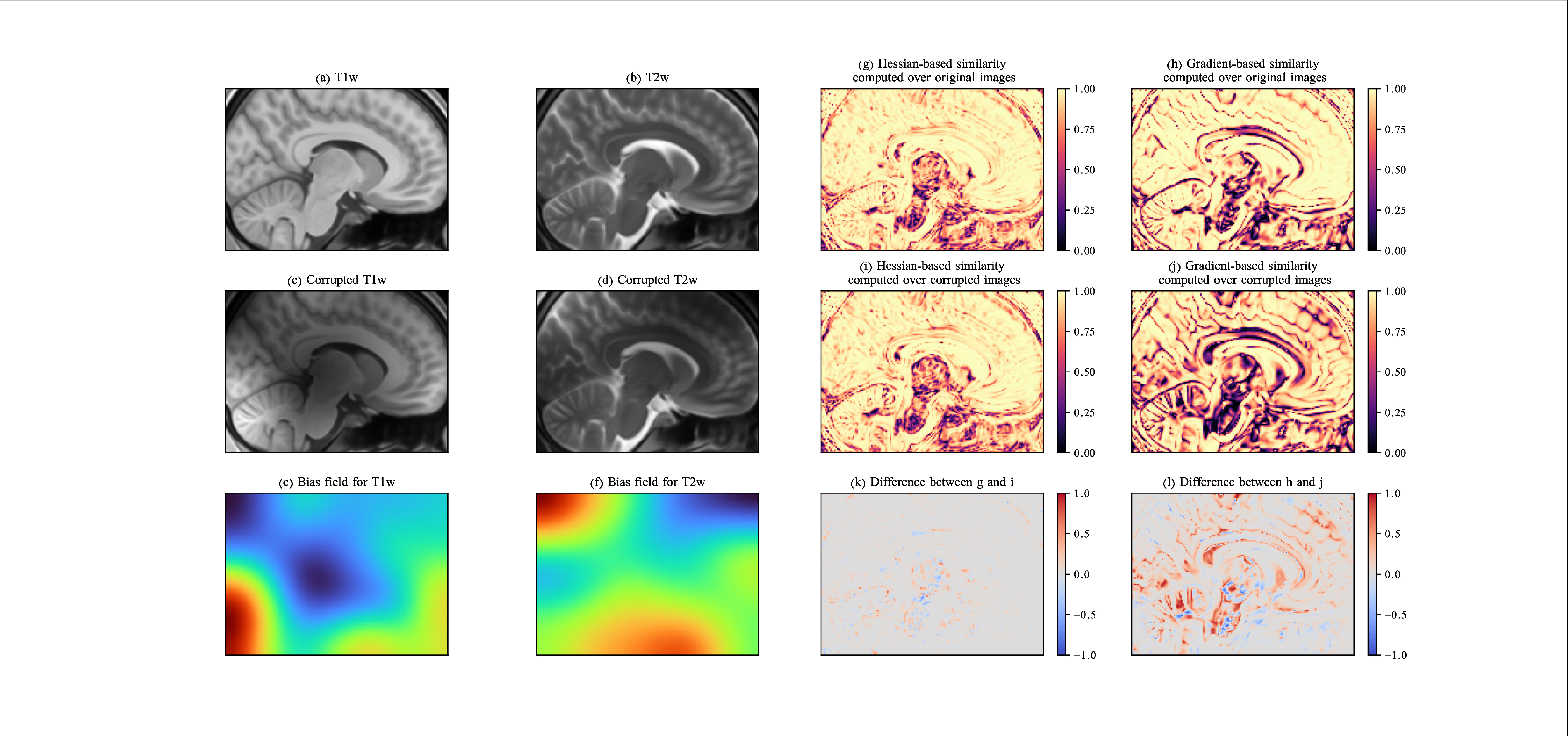}
\caption{The effect of intensity nonuniformities on the behavior of the Hessian-based similarity metric and gradient orientation alignment. (a) and (b) show T1w and T2w MNI-ICBM152 templates~\cite{ref27}. (c) and (d) are derived by multiplying (a) and (b) to (e) and (f) as synthetic bias fields, respectively. (g), (h), (i) and (j) show the computed Hessian-based similarity metric and gradient orientation alignment on original and corrupted image pairs. The difference between computed similarity maps are shown in (k) and (l).} \label{fig2}
\end{figure}

\section{Experiments}
\subsection{Robustness to Intensity Nonuniformities}
Intensity nonuniformities can degrade the precision of any analysis in medical image computing. Therefore, comping up with a similarity metric that is robust to these low-frequency artefacts is beneficial. To examine qualitatively and visually how robust our metric is to intensity nonuniformities, we compute the metric all over a pair of  registered images, before and after applying synthetic bias fields. Figure \ref{fig2} shows this process for T1w and T2w MNI-ICBM152 templates~\cite{ref27} on the saggital view. As it can be seen, the added artifacts have a negligible effect on the Hessian-based similarity metric; however, their effect on gradient orientation alignment is notable.

\subsection{Quantitative Results}
We test the registration performance of our method in the context of image-guided neurosurgery for aligning preoperative T1w MR volumes to preresection intraoperative ultrasound volumes of the BITE database~\cite{ref10}. The open-access BITE database contains scans of 14 patients with brain cancer, along with homologous manually annotated landmarks. We reconstruct ultrasound volumes from the ultrasound slices with a voxel size of \(0.5\times0.5\times0.5\) mm using IBIS~\cite{ref3}. We resample MR volumes to the same voxel size. We use provided landmarks to compute mean target registration error (mTRE) as a measure of accuracy.

To run our registration experiments, We set \(N\) to 5000 and the standard deviation for computing gaussian derivatives to 1.5~mm. Maximum displacement in each direction and maximum rotation around each axis is set to 10~mm and 5~degrees, respectively. Maximum shear and scaling in each direction are both set to 5\%. For optimization using Differential Evolution we use best/1/bin strategy~\cite{ref11} and set the population size and maximum number of iterations to 24 and 200, respectively. Crossover probability is set to \(0.7\) and differential weight will be randomly sampled from \((0.5,1)\) interval. Whenever the standard deviation of population’s cost falls below 0.2\% of the mean of population’s cost, the optimization will terminate.

As a baseline, we implement gradient orientation alignment and run it in the same framework that we have developed. We are choosing this method for two main reasons; first, its effectiveness for MR to ultrasound registration task has been previously demonstrated~\cite{ref12}, and second, it follows the same logic as our proposed metric.

\begin{figure}
\includegraphics[trim={-4cm 0cm -4cm 0cm},clip,width=\textwidth]{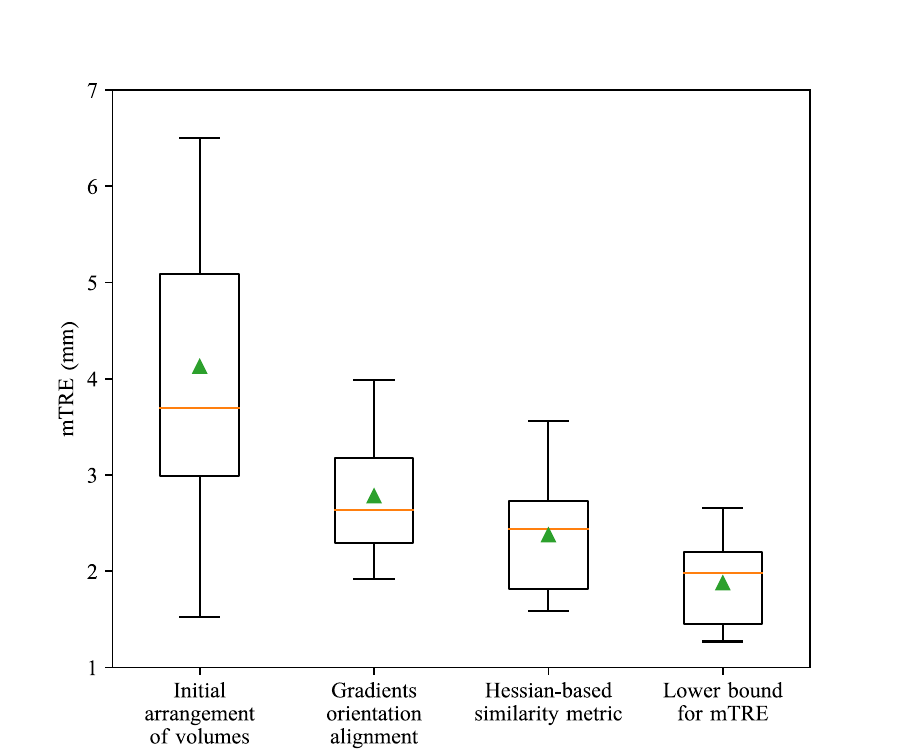}
\caption{Box plot of mTRE for registration of preoperative MR to pre-resection ultrasound on BITE database. First box on the left shows mTRE before registering MR to ultrasound. The box on the left shows the minimum mTRE that is achievable using an affine transformation based solely on the landmarks. Orange lines and green triangles represent the median and the mean, respectively.} \label{fig3}
\end{figure}

The results of the registration process are shown in Table \ref{tab1} and Figure \ref{fig3}. CPU time of each method for each case during preprocessing and optimization is reported, too. The preprocessing stage includes parsing images, random sampling, and computing image derivatives. All tests are performed on an Intel\textsuperscript{\circledR} Core\textsuperscript{TM} i7-11370H @ 3.30GHz CPU.

\begin{table}
\caption{Mean target registration error (mTRE) and CPU time for registration of preoperative MR to pre-resection ultrasound on BITE database}
\centering
\begin{tabular}{|c|c|c|c|c|c|c|}
\cline{2-7}
\multicolumn{1}{c|}{}& \multicolumn{3}{c|}{\textbf{Hessian-based similarity metric}} & \multicolumn{3}{c|}{\textbf{Gradient orientation alignment}} \\ \hline
\multirow{2}{8mm}{case no.} & \multirow{2}{1cm}{mTRE (mm)} & \multicolumn{2}{c|}{CPU time (s)} & \multirow{2}{1cm}{mTRE (mm)} & \multicolumn{2}{c|}{CPU time (s)} \\ \cline{3-4}\cline{6-7}
& & preprocess. & optim. & & preprocess. & optim. \\ \hline
2 & 1.78 (0.4-4.5) & 30.8 & 4.9 & 2.96 (1.1-5.7) & 13.1 & 3.2 \\
3 & 2.94 (0.7-5.8) & 48.8 & 5.7 & 2.55 (0.6-6.4) & 21.1 & 3.8 \\
4 & 1.63 (0.3-4.2) & 24.9 & 7.0 & 2.29 (0.9-4.5) & 10.7 & 4.3 \\
5 & 2.18 (0.4-6.6) & 33.9 & 3.0 & 2.14 (0.3-6.4) & 14.5 & 5.5 \\
6 & 1.82 (0.4-3.6) & 34.0 & 4.1 & 1.92 (0.8-3.5) & 14.4 & 3.4 \\
7 & 2.44 (0.8-5.0) & 40.2 & 5.0 & 3.17 (1.2-5.9) & 17.2 & 4.0 \\
8 & 2.71 (0.8-5.1) & 37.9 & 5.5 & 3.18 (0.5-6.2) & 16.2 & 3.1 \\
9 & 2.60 (0.5-5.8) & 27.2 & 4.8 & 2.64 (0.5-6.5) & 11.5 & 4.4 \\
10 & 1.84 (0.4-3.9) & 37.3 & 5.9 & 2.28 (0.7-5.7) & 16.6 & 4.0 \\
11 & 1.59 (0.6-3.2) & 32.9 & 5.6 & 2.57 (0.3-5.4) & 14.2 & 3.0 \\
12 & 3.06 (0.7-5.6) & 32.0 & 3.1 & 2.88 (0.7-6.0) & 13.6 & 3.2 \\
13 & 3.56 (0.9-6.3) & 28.3 & 4.6 & 3.99 (1.4-7.6) & 12.2 & 5.7 \\
14 & 2.73 (0.2-4.8) & 35.7 & 3.0 & 3.61 (2.0-6.5) & 15.6 & 3.4 \\ \hline
mean & 2.37 & 34.15 & 4.78 & 2.78 & 14.68 & 3.92 \\ \hline
\end{tabular}
\label{tab1}
\end{table}

\begin{figure}
\includegraphics[width=\textwidth]{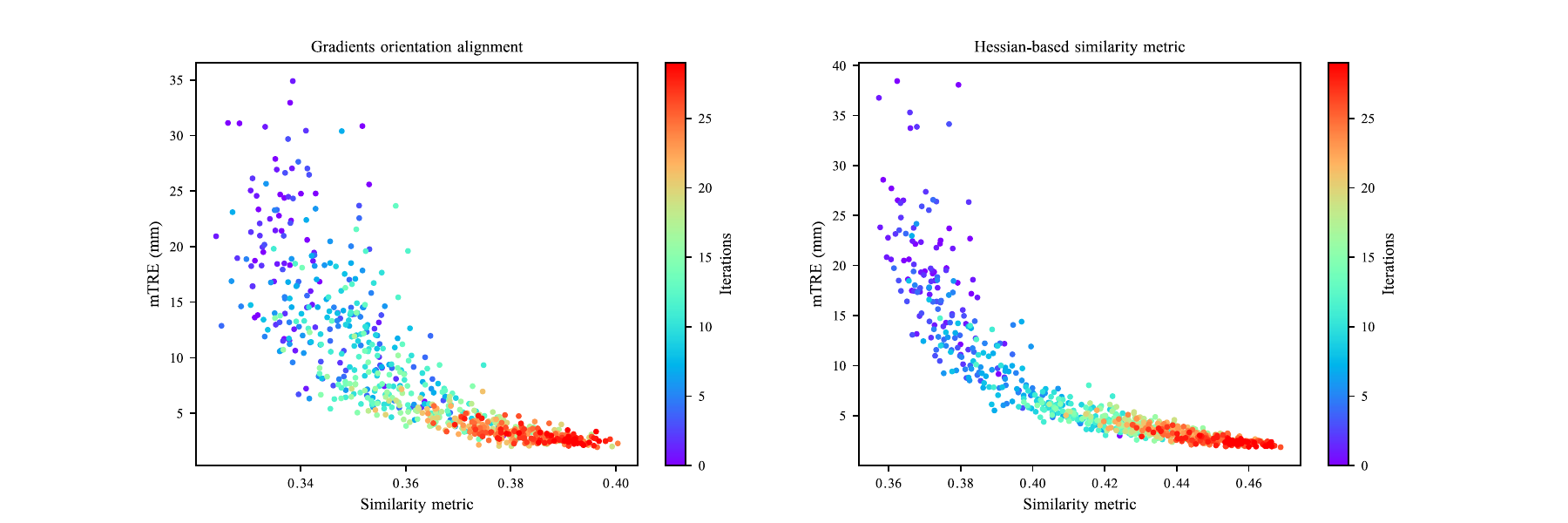}
\caption{Scatter plots of mTRE vs. similarity for deformations evaluated during the optimization for one MR-iUS pair. The color of each point shows the iteration in which the deformation was evaluated. We can see that the Hessian-based metric has less variability in optimization.} \label{fig4}
\end{figure}

To further monitor the behavior of each similarity metric, we plot the similarity score vs. mTRE for each deformation that is evaluated during the optimization. Figure~\ref{fig4} shows this scatter plot for case 10 of the BITE database. As it can be seen, both plots have the same general trend; as the similarity metric increases, mTRE decreases. In addition, the points derived from the Hessian-based similarity metric are spread out. This can imply a better capture range for this metric. Scatter plots of all other cases can be found in supplementary material.

We carry out a statistical test to determine whether the two metrics demonstrate statistically significant difference in registration accuracy. Our null hypothesis assumes the two methods result in the same distribution for mTRE. Our alternative hypothesis states that the Hessian-based similarity metric results in lower mTRE. Using a paired t-test, we achieve a p-value of 0.0048. Thus, the null hypothesis cannot be accepted, i.e., there is statistically significant improvement in the accuracy when our proposed similarity metric is used.

\section{Discussion and Conclusion}
In this paper, we introduce a new similarity metric for multimodal medical image registration. Our primary contribution has been deriving a closed form expression for computing the metric. We present geometrical insight into the mechanics of our proposed similarity metric. We address an efficient way for transforming the Hessian matrix to prevent recomputing it after applying deformations. The Hessian-based similarity metric is local, robust to intensity nonuniformities and computable at arbitrary scales. Moreover, the similarity metric is very flexible and can be used in a variety of registration tasks. The metric has shown a significant improvement in registering MR to ultrasound compared to a previously proposed similarity metric. However, this improvement comes at the cost of more computation, both in preprocessing and optimization stages. Since the preprocessing can be completed offline, its duration should not be a major concern. The increase in optimization time does not affect the usability of the method for online use cases.

For comparison, we list the results of some previously proposed methods on the BITE database. CoCoMI~\cite{ref16} and SeSaMI~\cite{ref24} have reported a mean error of 2.35~mm and 2.29~mm, respectively. Both of these methods use variants of mutual information as similarity metric along with cubic BSpline as deformation field. LC\textsuperscript{2}~\cite{ref25} method which was originally developoed for US-MR registration based on the physics of ultrasound, has reported a mean error of 2.49~mm using rigid transformation. In another work, through introducing a modality-independent binary descriptor called miLBP~\cite{ref26} for deformable registration, an error of 2.15~mm has been reported. With our Hessian-based similarity metric, the registration error is 2.37~mm, which is very close to that of other publications. It must be noted that unlike our method, other methods usually take advantage of nonlinear transformations, pyramid registration, and computationally-intensive preprocessing steps for outlier suppression or selecting regions with structures. We have acheived high accuracies using our proposed metric only with affine transformations. Therefore, we believe the Hessian-based similarity metric has great potentials for addressing multimodal registration tasks.

As future work, we plan to implement our proposed metric in a deformable registration scheme and test it on a broad range of modality pairs and registration tasks and compare its performance to that of the state-of-the-art methods.

%
%
%
%

\end{document}